\documentclass{elsart}

\usepackage{graphicx}
\usepackage{amssymb}
\usepackage{color}

\begin{document}

\begin{frontmatter}

\title{Evolution of ferromagnetic order in URhGe alloyed with Ru, Co and Si}

\author[Delft]{S. Sakarya},
\author[Amsterdam]{N. T. Huy},
\author[Delft]{N. H. van Dijk},
\author[Amsterdam]{A. de Visser\corauthref{cor}},
\corauth[cor]{Corresponding author.} \ead{devisser@science.uva.nl}
\author[Delft]{M. Wagemaker},
\author[Amsterdam]{A. C. Moleman},
\author[Amsterdam]{T. J. Gortenmulder},
\author[Amsterdam]{J. C. P. Klaasse},
\author[Karlsruhe1]{M. Uhlarz},
\author[Karlsruhe1,Karlsruhe2]{H. v. L\"{o}hneysen}

\address[Delft]{Department of Radiation, Radionuclides \& Reactors, Delft University of Technology,
Mekelweg 15, 2629 JB Delft, The Netherlands}
\address[Amsterdam]{Van der Waals - Zeeman Institute, University of Amsterdam, Valckenierstraat~65, 1018 XE
Amsterdam, The Netherlands}
\address[Karlsruhe1]{Physikalisches Institut, Universit\"{a}t Karlsruhe, D-76128 Karlsruhe, Germany}
\address[Karlsruhe2]{Forschungszentrum Karlsruhe, Institut f\"{u}r Festk\"{o}rperphysik, D-76021 Karlsruhe,
Germany}

\begin{abstract}

We have investigated the evolution of ferromagnetic order in the
correlated metal URhGe (Curie temperature $T_{\rm C} = $9.5~K) by
chemical substitution of Ru, Co and Si. Polycrystalline samples
URh$_{1-x}$Ru$_x$Ge ($x \leq $0.6), URh$_{1-x}$Co$_x$Ge ($x \leq
$0.9) and URhGe$_{1-x}$Si$_x$ ($x \leq $0.2) have been prepared
and the magnetic properties have been investigated by
magnetization and transport experiments. In the case of Ru doping,
$T_{\rm C}$ initially increases, but then decreases linearly as a
function of $x$ and is completely suppressed for $x_{\rm cr}
\approx 0.38$. The Curie temperature in the URh$_{1-x}$Co$_x$Ge
series has a broad maximum $T_{\rm C} = 20$~K near $x=0.6$ and
then drops to 8 K for $x=0.9$. In the case of Si doping $T_{\rm
C}$ stays roughly constant. We conclude that the alloy systems
URh$_{1-x}$Ru$_x$Ge and URh$_{1-x}$Co$_x$Ge are interesting
candidates to study the ferromagnetic instability.

\end{abstract}
\begin{keyword}
ferromagnetism, URhGe, chemical substitution, quantum phase
transition \PACS 75.20.En \sep 75.30.Kz \sep 75.30.Mb
\end{keyword}
\end{frontmatter}

\section{Introduction}

Recently, URhGe has attracted significant attention because
ferromagnetism (Curie temperature $T_{\rm C} = 9.5$~K) and
unconventional superconductivity ($T_{\rm s} = 0.25$~K) coexist at
ambient pressure~\cite{Aoki-Nature-2001}. The superconducting
state is believed to have its origin in the proximity to a
ferromagnetic instability. Near the quantum critical point (QCP),
which can be reached by tuning $T_{\rm C}$ to 0 K, enhanced
ferromagnetic spin fluctuations mediate Cooper pairing (of the
spin-triplet type~\cite{Hardy-PRL-2005}). The important role of
critical magnetic fluctuations in URhGe is furthermore indicated
by field-induced superconductivity for a magnetic field $B$
directed along the orthorhombic $b$-axis~\cite{Levy-Science-2005}.
It has been suggested that the high-field superconducting phase is
due to magnetic fluctuations associated with the spin
reorientation process which takes place at $B \approx
12~$T~\cite{Levy-Science-2005}. Clearly, it is of considerable
interest to further investigate the magnetic properties of URhGe,
especially with respect to the proximity to a magnetic
instability.

The crystallographic, magnetic, transport and thermal properties
of URhGe have been investigated in much detail on polycrystalline
as well as on single-crystalline samples
~\cite{Troc-JMMM-1988,Chevalier-JMMM-1988,Tran-JMMM-1990,Buschow-JAP-1990,DeBoer-PhysicaB-1990,Tran-JMMM-1998,Hagmusa-PhysicaB-2000,Prokes-PhysicaB-2002,Sakarya-PRB-2003}.
URhGe crystallizes in the TiNiSi structure (space group $Pnma$)
~\cite{Chevalier-JMMM-1988}. Itinerant ferromagnetic order is
found below $T_{\rm C} = 9.5$~K \cite{Troc-JMMM-1988} and the
ordered moment of about $0.4 \mu_{\rm B}$/U-atom points along the
orthorhombic $c$-axis
~\cite{Aoki-Nature-2001,Prokes-PhysicaB-2002}. The linear
electronic coefficient in the specific heat $\gamma = 160$
mJ/molK$^2$ is enhanced, which indicates that URhGe is a
correlated metal~\cite{Buschow-JAP-1990}.

Here we report the evolution of ferromagnetic order in URhGe by
doping with Ru or Co on the Rh site and Si on the Ge site. Our
choice for the elements Ru and Co is motivated by the fact that
URuGe and UCoGe are isostructural to URhGe and have a paramagnetic
ground state
\cite{Troc-JMMM-1988,Buschow-JAP-1990,Canepa-JAC-1996}. This
indicates doping with Ru or Co could possibly lead to a reduction
of $T_{\rm C}$ and the approach to the magnetic instability. A
large difference between both dopants is that Co is an
isoelectronic substitution, while Ru is not. Notice that all other
neighbouring isostructural UTGe compounds (here T is a transition
metal) have a magnetic ground state: UIrGe, UNiGe and UPtGe are
antiferromagnets, while UPdGe is a ferromagnet
~\cite{Troc-JMMM-1988,Sechovsky-Handbook-1998}. The compound URhSi
is also isostructural to URhGe and ferromagnetic with a Curie
temperature of $9.5$~K
\cite{Troc-JMMM-1988,Buschow-JAP-1990,Tran-JMMM-1998}.

The magnetic properties were studied by magnetization measurements
on polycrystalline samples URh$_{1-x}$Ru$_x$Ge with $x \leq 0.6$,
URh$_{1-x}$Co$_x$Ge with $x \leq 0.9$ and URhGe$_{1-x}$Si$_x$ with
$x \leq 0.2$. The URh$_{1-x}$Ru$_x$Ge alloys were also
investigated by electrical resistivity experiments. In the case of
Ru doping $T_{\rm C}$, after an initial weak increase, decreases
linearly and vanishes at a critical Ru concentration $x_{\rm cr}
\approx 0.38$. Co doping leads to an increase of $T_{\rm C}$ up to
20~K for $x = 0.60$, beyond which $T_{\rm C}$ drops to 8 K for
$x=0.9$. In the URhGe$_{1-x}$Si$_x$ system no significant change
of $T_{\rm C}$ was observed up to $x = 0.20$. A preliminary
account of the evolution of magnetism in the URh$_{1-x}$Ru$_x$Ge
series has been reported in Ref.~\cite{Sakarya-PhysicaB-2006}.

\section{Sample preparation, X-ray diffraction and experimental techniques}

Polycrystalline samples URh$_{1-x}$Ru$_x$Ge ($x \leq 0.6$),
URh$_{1-x}$Co$_x$Ge ($x \leq 0.9$) and URhGe$_{1-x}$Si$_x$ ($x
\leq 0.2$) were prepared by arc melting the constituents U, Rh,
Ru, Co (all 3N purity) and Ge and Si (both 5N purity) under a
high-purity argon atmosphere. Each sample was melted several times
and turned over after each melt to improve the homogeneity. The
as-cast buttons were wrapped in Ta foil and annealed in quartz
tubes under high vacuum for ten days at 875~$^\circ$C. Samples for
magnetization and transport experiments were cut by spark-erosion.
Electron probe micro analysis showed the single phase nature of
the samples within the resolution of 2\%. X-ray powder diffraction
confirmed the orthorhombic TiNiSi structure (space group $Pnma$)
\cite{Prokes-PhysicaB-2002,Lloret-PhDthesis-1988} for all samples.

\begin{figure} \center
\resizebox{!}{6cm}{\includegraphics{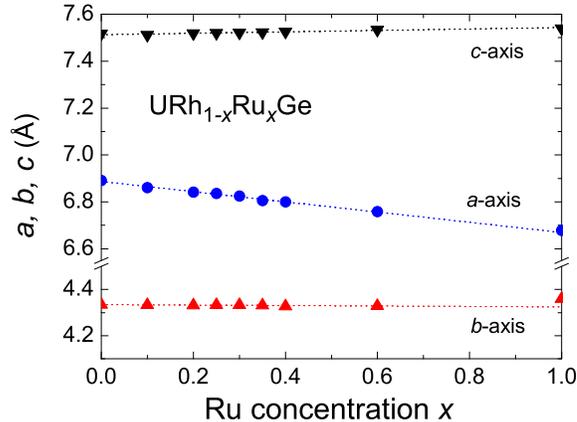}}
\caption{\label{Figure_URhRuGe_Lattice_parameters}Lattice
parameters $a$ $(\bullet)$, $b$ $(\blacktriangle)$, and $c$
$(\blacktriangledown)$ of URh$_{1-x}$Ru$_x$Ge as a function of the
Ru concentration $x$ measured at room temperature. Data for URuGe
are taken from Ref.\cite{Buschow-JAP-1990}}.
\end{figure}

The lattice parameters of the URh$_{1-x}$Ru$_x$Ge series have been
determined by X-ray diffraction for samples with $x \leq 0.60$.
The results are shown in
figure~\ref{Figure_URhRuGe_Lattice_parameters} together with
literature data for pure URuGe~\cite{Buschow-JAP-1990}. For URhGe
we obtain $a = 6.887$~\AA, $b = 4.334~$\AA ~and $c = 7.513~$\AA
~in good agreement with literature values (the uncertainty in the
determination of the lattice parameters is about 0.1~\%). The
variation of the lattice parameters upon doping is anisotropic.
The $a$ lattice parameter shows the largest variation, it reduces
linearly with increasing $x$. The $c$ parameter shows a small
increases, while the $b$ parameter remains almost constant. The
unit cell volume $\Omega$ = 224.2~\AA$^3$ for URhGe follows
Vegard's law and decreases linearly at a rate of 0.067~\AA$^3$ per
at.\% Ru. The extrapolated value of $\Omega$ for URuGe amounts to
217.5~\AA$^3$, which is slightly smaller than the literature value
219.5~\AA$^3$~\cite{Buschow-JAP-1990}. This difference is mainly
due to the smaller extrapolated value for the $b$ lattice
parameter compared to the literature value (see
figure~\ref{Figure_URhRuGe_Lattice_parameters}).

%Figure_URhRuGe_Lattice_parameters

\begin{figure} \center
\resizebox{!}{6cm}{\includegraphics{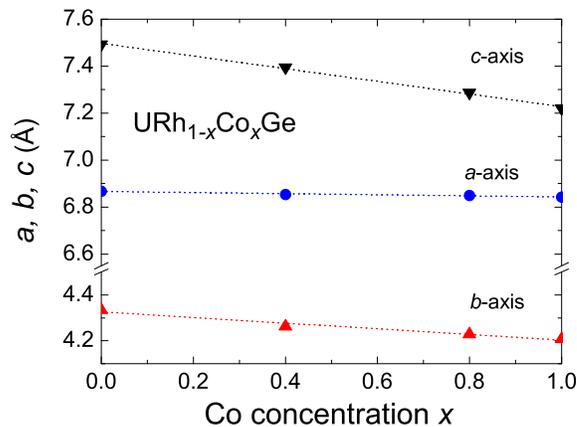}}
\caption{\label{Figure_URhCoGe_Lattice_parameters}Lattice
parameters $a$ $(\bullet)$, $b$ $(\blacktriangle)$, and $c$
$(\blacktriangledown)$ of URh$_{1-x}$Co$_x$Ge as a function of the
Co concentration $x$ measured at room temperature.}
\end{figure}

The measured variation of the lattice parameters in the
URh$_{1-x}$Co$_x$Ge series is shown in
figure~\ref{Figure_URhCoGe_Lattice_parameters}. Here the $b$ and
$c$ lattice parameter show a linear decrease, while the $a$
parameter remains almost constant. The unit cell volume decreases
linearly at a rate of 0.152~\AA$^3$ per at.\% Co. For UCoGe we
obtain $a = 6.845$~\AA, $b = 4.206~$\AA ~and $c = 7.222~$\AA, with
$\Omega= 207.95$~\AA$^3$, in good agreement with the literature
values \cite{Canepa-JAC-1996}.

%Figure_URhCoGe_Lattice_parameters

The lattice parameters for the URhGe$_{1-x}$Si$_x$ alloys
($x=0.10$ and $x=0.20$) have not been determined. However, given
the literature values for URhSi ($a = 7.024$~\AA, $b = 4.121~$\AA
~and $c = 7.458~$\AA) \cite{Prokes-JAP-1996} and assuming Vegard's
law, we conclude that the $a$ parameter expands and the $b$ and
$c$ parameters contract with increasing Si content. The unit cell
volume decreases linearly at a rate of 0.084~\AA$^3$ per at.\% Si
to $\Omega = 215.9$~\AA$^3$ for URhSi \cite{Prokes-JAP-1996}.

The dc magnetization $M(T,B)$ was measured in a Quantum Design
SQUID magnetometer. Temperature scans were made between 1.8 and
20~K in a field of 0.01~T and between 2 and 300~K in a field of
1~T. In both cases the data were taken after field cooling. Field
scans of the magnetization were made in fields up to 5.5~T at
several temperatures. The electrical resistivity, $\rho(T)$, was
measured using a standard four probe low-frequency ac-technique in
zero magnetic field from 2 to 300~K in a MagLab system (Oxford
Instruments).

\section{Experimental results}

\subsection{U(Rh,Ru)Ge alloys}

\begin{figure} \center
\resizebox{!}{8cm}{\includegraphics{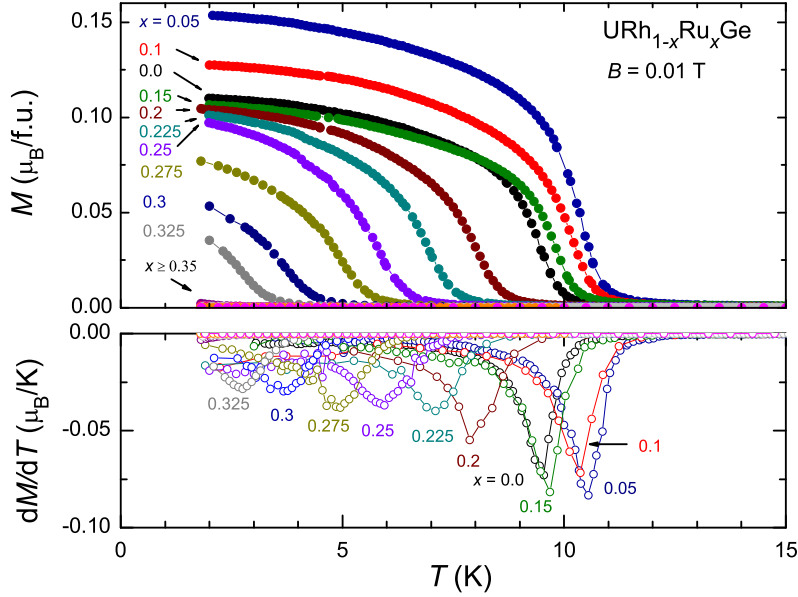}}
\caption{\label{Fig_URhRuGe_Magnetization_Ru}Upper frame:
Temperature variation of the dc magnetization measured in a field
$B = 0.01$~T of URh$_{1-x}$Ru$_x$Ge alloys with $x \leq 0.6 $ as
indicated. Notice $T_{\rm C}$ first increases and has a maximum
value for $x= 0.05$. For $0.35 \leq x \leq 0.6$ magnetic order is
not observed above $T=1.8$ K. Lower frame: Temperature derivative
of the magnetization.}
\end{figure}

The temperature variation of the magnetization, $M(T)$, of the
URh$_{1-x}$Ru$_x$Ge series measured in a field of 0.01~T is shown
in figure~\ref{Fig_URhRuGe_Magnetization_Ru}. Also shown, in the
lower part of the figure, is the derivative ${\rm d}M(T) / {\rm
d}T$. The inflection point in $M(T)$ or the temperature at which
${\rm d}M(T) / {\rm d}T$ has a minimum defines the Curie
temperature $T_{\rm C}$. For pure URhGe we find $T_{\rm C} =
9.6$~K, in good agreement with previous values reported in the
literature
\cite{Aoki-Nature-2001,Troc-JMMM-1988,Buschow-JAP-1990}. Upon
replacing Rh by Ru the ferromagnetic transition first shifts
upwards to 10.6~K for $x=0.05$. For higher concentrations magnetic
order is suppressed in a monotonic way. At $x=0.15$, $T_{\rm C}$
attains the same value as for pure URhGe and beyond $x=0.20$
$T_{C}$ decreases approximately linearly with $x$ at a rate of
$0.45$~K/at.\% Ru. For the samples with $x=0.35, 0.375, 0.40$, and
0.60 $T_{\rm C}$ no magnetic transition was observed in the
measured temperature range ($T > 1.8$~K).

%Fig_URhRuGe_Magnetization_Ru

\begin{figure} \center
\resizebox{!}{7cm}{\includegraphics{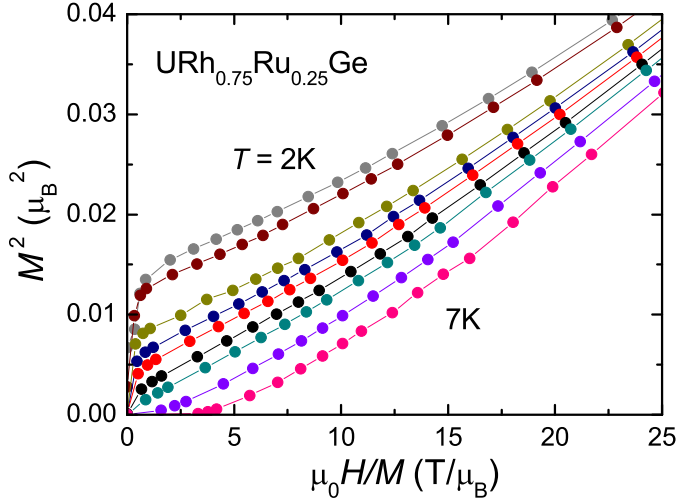}}
\caption{\label{Fig_URhRuGe_Arrott_Ru}Arrott plot of the
magnetization of URh$_{0.75}$Ru$_{0.25}$Ge. The isotherms  were
measured (from top to down) at $T =$ 2.0 , 3.0, 4.5, 5.0, 5.3,
5.7, 6.0, 6.5 and 7.0 K. The isotherm through the origin
determines $T_{\rm C} = 6.0$~K.}
\end{figure}

For all samples in addition the field variation of the
magnetization, $M(B)$, was measured up to 5.5~T at a number of
fixed temperatures. By making Arrott plots, i.e. by plotting the
data as $M^2$ versus $\mu_0 H/M$, we identify $T_{\rm C}$ by the
isotherm that intersects the origin. A typical Arrott plot is
presented in figure~\ref{Fig_URhRuGe_Arrott_Ru} for $x=0.25$.
Ideally the isotherms should be linear. For the
URh$_{1-x}$Ru$_x$Ge alloys, the initial increase of $M^2$ as a
function of $\mu_0 H/M$ is caused by demagnetization effects. The
upward curvature at higher values of $\mu_0 H/M$ is due to
reorientation processes (from the easy axis to the applied
magnetic field) ~\cite{Levy-Science-2005} of the magnetic moments
in our polycrystalline samples. The Curie temperatures deduced
from the Arrott plots (neglecting the small error in the
determination of $T_{\rm C}$ due to demagnetization effects) are
in good agreement with those derived from the minimum in ${\rm
d}M(T) / {\rm d}T$. For $x=0.25$ we find $T_{\rm C} =6.0$ K. The
Arrott plot of the compound with $x=0.35$ (not shown) suggests
that ferromagnetism sets in near $T_{\rm C} \sim 1.3$~K. This
value of $T_{\rm C}$ is estimated by extrapolating the
intersection points of the isotherms with the $\mu_0 H / M $ axis
to the origin of the Arrott plot. For $x=0.375$ $T_{\rm C}$ is
close to zero, while for $x=0.40$ and $0.60$ the Arrott plots
clearly indicate a paramagnetic ground state.

%Fig_URhRuGe_Arrott_Ru

\begin{figure}
\center
\resizebox{!}{7cm}{\includegraphics{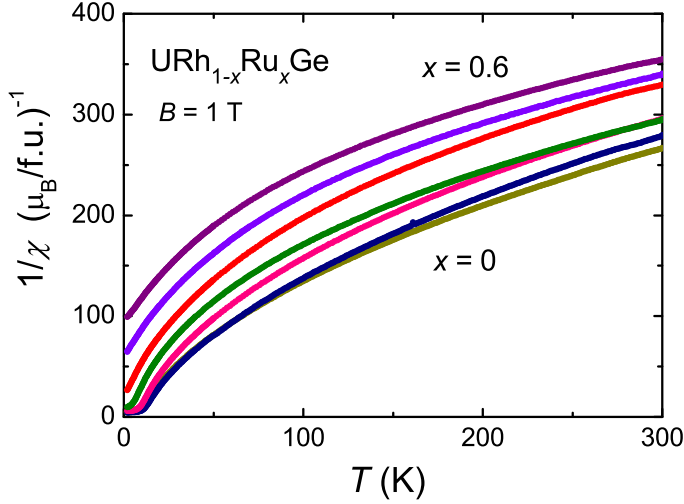}}
\caption{\label{Fig_URhRuGe_Susceptibility_Ru}Temperature
variation of the inverse susceptibility $1 / \chi$ of
URh$_{1-x}$Ru$_x$Ge alloys measured in a field of 1~T. Ru
concentrations are (from bottom to top) $x =$ 0, 0.1, 0.2, 0.3,
0.4, 0.5 and 0.6.}
\end{figure}

In figure~\ref{Fig_URhRuGe_Susceptibility_Ru} we have plotted the
reciprocal susceptibility, $1/ \chi$, of a few selected
URh$_{1-x}$Ru$_x$Ge alloys, measured in a field $B=1$ T in the
temperature range 2-300 K. The strong magneto-crystalline
anisotropy, observed on single-crystalline
samples~\cite{Prokes-PhysicaB-2002}, hampers a proper analysis of
the high-temperature local-moment susceptibility in our
polycrystalline samples, using the modified Curie-Weiss law $\chi
(T) = C / (T - \theta) + \chi_0$ (here $\chi_0$ represents a
temperature independent contribution). However, the overall upward
shift of the curves with increasing Ru contents indicates an
increasing (antiferromagnetic) interaction strength $\theta$. The
analysis is further complicated by the strong curvature of $1/
\chi$ versus $T$ which signals crystalline electric field effects.
Note that in pure URhGe the easy-axis ($c$-axis) susceptibility
measured on a single crystal does follow a modified Curie Weiss
behavior with $\theta \approx T_{\rm C}$, as demonstrated in
Ref.~\cite{Aoki-Nature-2001}.

%Fig_URhRuGe_Susceptibility_Ru

\begin{figure} \center
\resizebox{11.9cm}{!}{\includegraphics{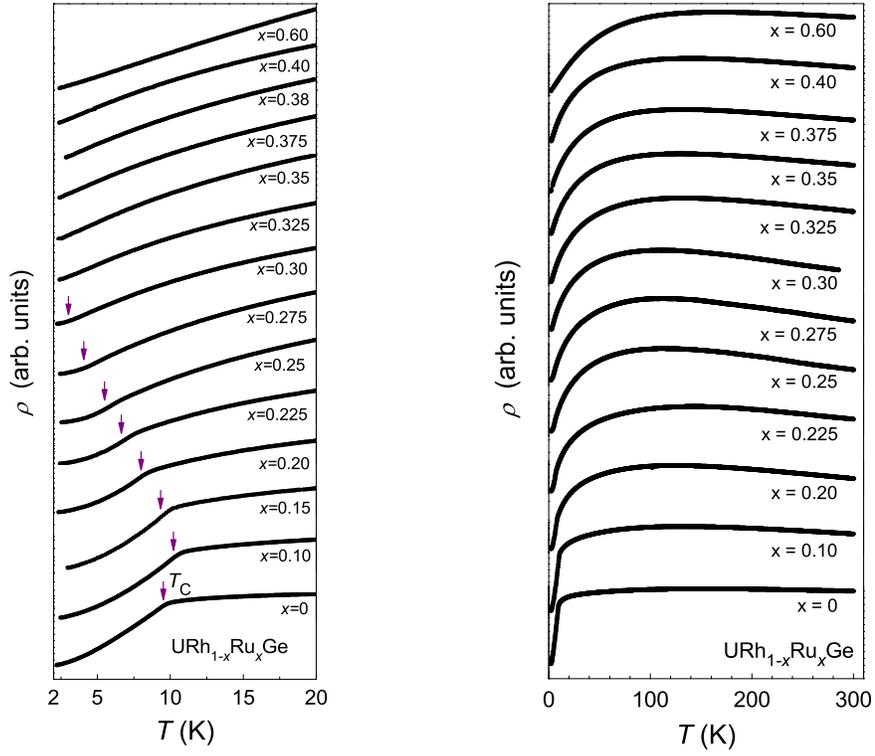}}
\caption{\label{Fig_URhRuGe_Resistivity_Ru}Temperature dependence
of the electrical resistivity $\rho$ in arbitrary units of
URh$_{1-x}$Ru$_x$Ge alloys for $0 \leq x \leq 0.6$ as indicated.
Left panel: $2$~K $\leq T \leq 20$~K. The Curie temperatures are
indicated by arrows. Right panel: $2$~K $\leq T \leq 300$~K.}
\end{figure}

The electrical resistivity $\rho (T)$ of the URh$_{1-x}$Ru$_{x}$Ge
alloys is shown in figure~\ref{Fig_URhRuGe_Resistivity_Ru}. Note
the vertical scale is in arbitrary units. For $x \geq 0.60$ the
overall temperature variation (see right panel in
figure~\ref{Fig_URhRuGe_Resistivity_Ru}) is consistent with the
formation of a Kondo-lattice, i.e. an increase of the resistivity
upon lowering $T$ below 300 K, a weak maximum in the temperature
range $70-130$ K and a steady drop signaling coherence at low
temperatures. For all doped samples the absolute variation of the
resistivity in the temperature interval $2-300$ K amounts to
150-250$~\mu \Omega$cm, which are usual values for uranium
intermetallics \cite{Tran-JMMM-1990,Sechovsky-Handbook-1998}. The
residual resistivity values $\rho_{0}$ are large (200-300$~\mu
\Omega$cm) and do not follow a systematic variation with Ru
concentration. This we mainly attribute to the brittleness of the
samples (cracks). Consequently, the residual resistance ratio's
(RRR = $R (300$K)/R($2$K)) are small ($\approx 2$). The left panel
in figure~\ref{Fig_URhRuGe_Resistivity_Ru} shows $\rho (T)$ in the
temperature interval 2-20 K. For $x=0$ the kink in $\rho (T)$
signals the Curie temperature, $T_{\rm C}= 9.4$~K, in agreement
with the magnetization data. Below $T_{\rm C}$ the resistivity is
dominated by spin-wave scattering, while for $T \geq T_{\rm C}$
spin-disorder scattering is dominant. With increasing $x$ the kink
becomes less pronounced. However, for all $x \leq 0.30$ $T_{\rm
C}$ can be identified by the maximum in d$\rho$/d$T$ (arrows in
figure~\ref{Fig_URhRuGe_Resistivity_Ru}). The Curie temperatures
determined in this way are in good agreement with those obtained
by the magnetization measurements.

%Fig_URhRuGe_Resistivity_Ru

\subsection{U(Rh,Co)Ge alloys}

%Fig_URhRuGe_Magnetization_Co
\begin{figure}
\center
\resizebox{!}{7cm}{\includegraphics{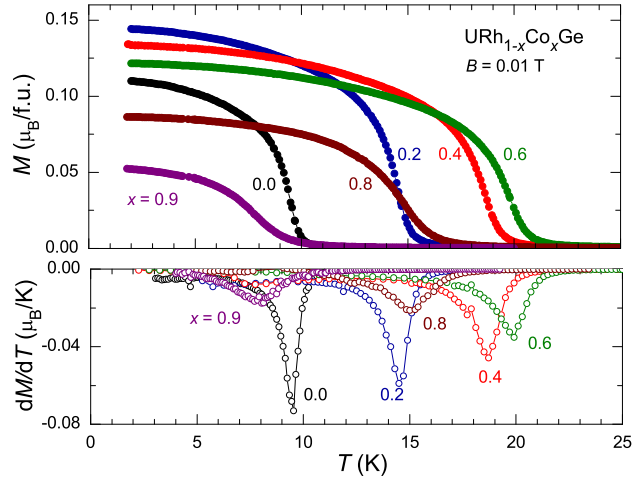}}
\caption{\label{Fig_URhRuGe_Magnetization_Co}Upper frame:
Temperature variation of the dc magnetization measured in a field
$B = 0.01$~T of URh$_{1-x}$Co$_x$Ge alloys with $x \leq 0.9 $ as
indicated. Notice $T_{\rm C}$ has a maximum value for $x= 0.6$.
Lower frame: Temperature derivative of the magnetization.}
\end{figure}

%Fig_URhRuGe_Arrott_Co
\begin{figure}
\center
\resizebox{!}{7cm}{\includegraphics{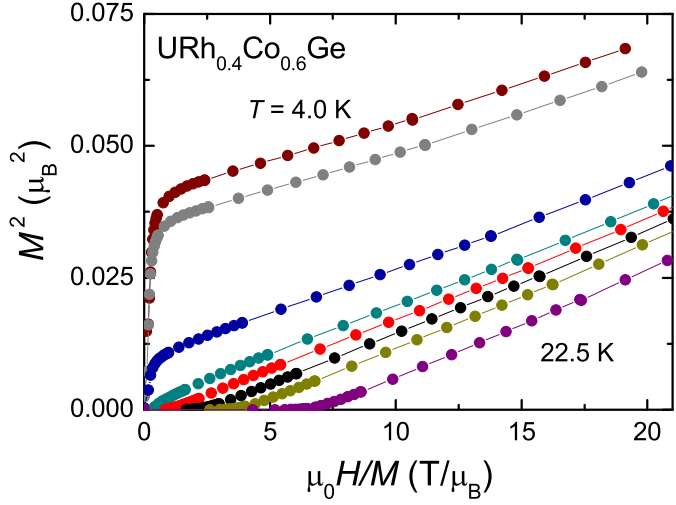}}
\caption{\label{Fig_URhRuGe_Arrott_Co}Arrott plot of the
magnetization of URh$_{0.4}$Co$_{0.6}$Ge. The isotherms  were
measured (from top to down) at $T =$ 4.0 , 10.0, 18.5, 20.0, 20.5,
21.0, 21.5 and 22.5 K. The isotherm through the origin determines
$T_{\rm C} = 20.0$~K.}
\end{figure}

\begin{figure} \center
\resizebox{!}{7cm}{\includegraphics{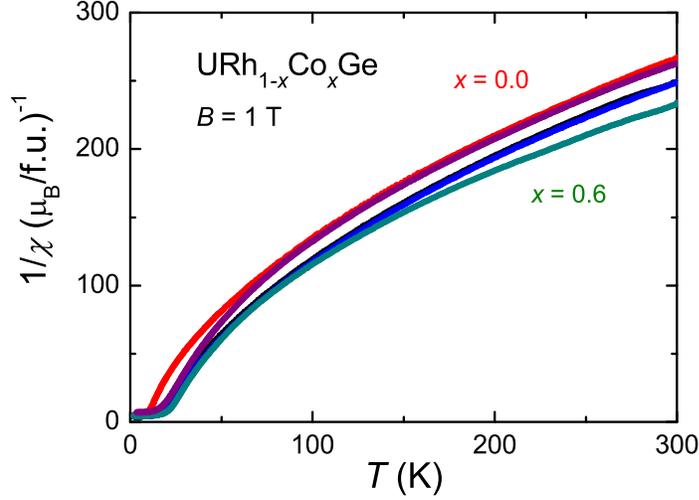}}
\caption{\label{Fig_URhRuGe_Susceptibility_Co}Temperature
variation of the inverse susceptibility $1 / \chi$ of
URh$_{1-x}$Co$_x$Ge alloys for measured in a field of 1~T. The Co
concentrations are $x =$ 0.0, 0.2, 0.4, 0.6, 0.8. The data for $x
= 0.2$ and 0.4, and $x= 0.0$ and 0.8 largely overlap.}
\end{figure}

The temperature variation of the magnetization, $M(T)$, and its
derivative ${\rm d}M(T) / {\rm d}T$, for the URh$_{1-x}$Co$_x$Ge
series measured in a field of 0.01~T is shown in
figure~\ref{Fig_URhRuGe_Magnetization_Co}. The effect of Co doping
differs from that by Ru doping. $T_{\rm C}$ increases
monotonically up to $x=0.6$, where it reaches a value of 20~K,
i.e. more than twice the value for $x=0$. For higher values of
$x$, $T_{\rm C}$ decreases and drops to $8.0$ K for $x=0.9$. Since
UCoGe has been reported to be paramagnetic (at least for $T \geq
1.2$~K) \cite{Troc-JMMM-1988,Buschow-JAP-1990}, the data suggest
that UCoGe is close to ferromagnetic order. The proximity to a
ferromagnetic instability was previously proposed on the basis of
high-field magnetization measurements (at $T=$ 4.2~K) in which a
relatively large field-induced magnetic moment of 0.58 $\mu_{B}$
was observed in a field of 35 T \cite{DeBoer-PhysicaB-1990}. The
variation of $T_{\rm C}$ with Co concentration has also been
tracked with help of Arrott plots. The Arrott plot for $x=0.6$,
where ferromagnetic order is most robust, is shown in
figure~\ref{Fig_URhRuGe_Arrott_Co}. In
figure~\ref{Fig_URhRuGe_Susceptibility_Co} we show the reciprocal
susceptibility measured in 1~T. The curves for various amounts of
Co doping are very similar, which indicates that the Curie-Weiss
constant $\theta$ does not vary much. The data for $x = 0.2$ and
0.4, and $x= 0.0$ and 0.8 largely overlap. This indicates the data
do not represent a polycrystalline average, and our samples
contain crystallites with preferred orientations.

%Fig_URhRuGe_Susceptibility_Co

\subsection{URh(Ge,Si) alloys}

%Fig_URhRuGe_Magnetization_Si
\begin{figure}
\center
\resizebox{!}{7cm}{\includegraphics{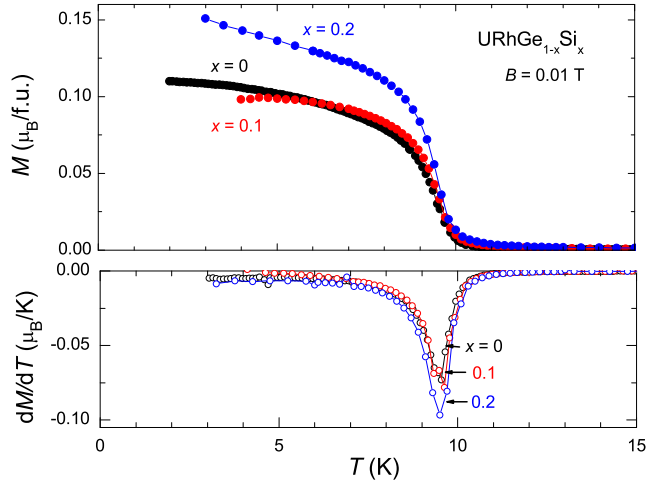}}
\caption{\label{Fig_URhRuGe_Magnetization_Si}Upper frame:
Temperature variation of the dc magnetization measured in a field
$B = 0.01$~T of URhGe$_{1-x}$Si$_x$ alloys with $x = 0$, 0.1 and
0.2 as indicated. Notice $T_{\rm C}$ stays roughly constant. Lower
frame: Temperature derivative of the magnetization.}
\end{figure}

In figure~\ref{Fig_URhRuGe_Magnetization_Si} the magnetization
measured in 0.01~T as a function of temperature is shown for the
URhGe$_{1-x}$Si$_x$ series with $x \leq 0.20$. Ferromagnetic order
is robust in the case of Si doping. The Curie temperatures deduced
from $\left({\rm d}M(T) / {\rm d}T\right)_{\rm min}$ agree with
those deduced from the Arrott plots, as expected. $T_{\rm C}$ does
not change with the Si content up to $x=0.20$. The reciprocal
susceptibility, plotted in
figure~\ref{Fig_URhRuGe_Susceptibility_Co}, shows only a weak
variation with the Si content.

%Fig_URhRuGe_Susceptibility_Si
\begin{figure}
\center
\resizebox{!}{7cm}{\includegraphics{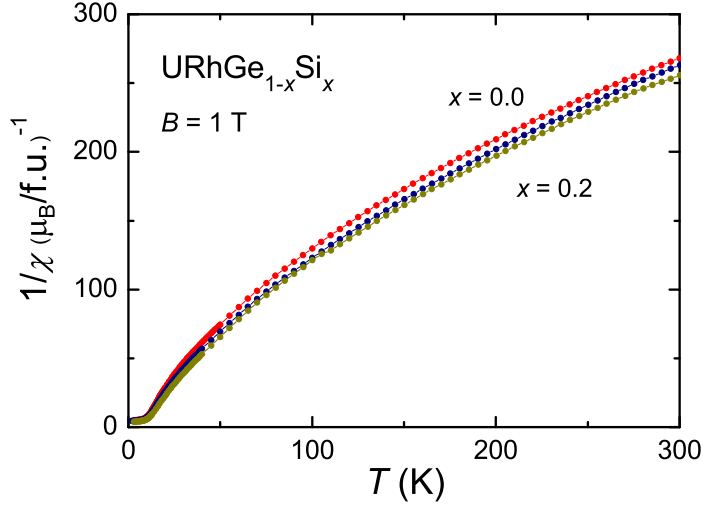}}
\caption{\label{Fig_URhRuGe_Susceptibility_Si}Temperature
variation of the inverse susceptibility $1 / \chi$ of
URhGe$_{1-x}$Si$_x$ alloys measured in a field of 1~T. Si
concentrations are (from top to bottom) $x =$ 0, 0.1 and 0.2.}
\end{figure}

\section{Discussion}

\begin{figure}
\center
\resizebox{!}{7cm}{\includegraphics{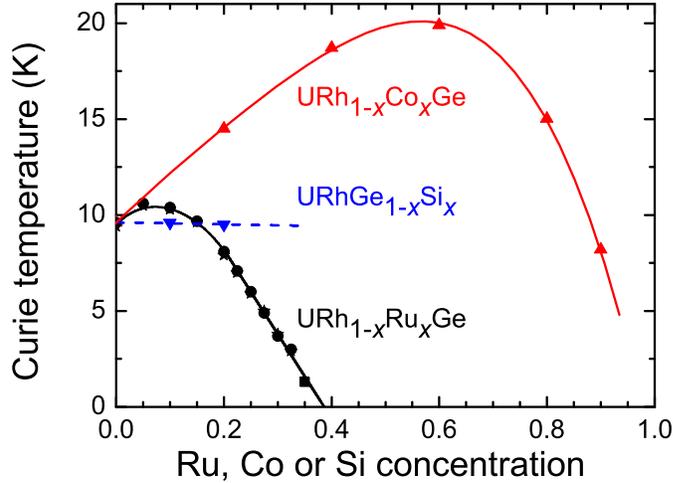}}
\caption{\label{Fig_URhRuGe_TC_versus_doping} Variation of the
Curie temperature $T_{\rm C}$ of URhGe doped with Ru ($\bullet$
from magnetization, $\bigstar$ from transport), Co
($\blacktriangle$) and Si ($\blacktriangledown$). The solid lines
serve as a guide to the eye. The data point for
URh$_{0.65}$Ru$_{0.35}$Si $(\blacksquare)$ is determined by
extrapolation in the Arrott plot for $x=$0.35 (see text). The
critical concentration for the suppression of ferromagnetic order
is $x_{\rm cr} \approx 0.38$ for Ru doping and $x_{\rm cr} \approx
1.0$ for Co doping.}
\end{figure}

The main results of our study of the evolution of ferromagnetic
order in URhGe doped with Ru, Co and Si, namely the variation
$T_{\rm C}(x)$, is presented in
figure~\ref{Fig_URhRuGe_TC_versus_doping}. The $T_{\rm C}(x)$
curves for Ru and Co substitution follow a similar trend: $T_{\rm
C}$ first increases, has a maximum near $x=0.05$ and $x=0.60$ for
Ru and Co doping, respectively, and then vanishes near $x=0.38$
and $x \sim 1.0$, respectively. However, in a quantitative measure
Ru is more than twice as effective as Co in suppressing $T_{\rm
C}$. Doping up to 20 at.\% Si on the Ge site does not suppress
ferromagnetic order and $T_{\rm C}$ remains 9.4 K. Since for the
end compound URhSi the Curie temperature has the same value
\cite{Troc-JMMM-1988}, the data suggest $T_{\rm C} \approx 9.4$~K
in the entire URh(Ge,Si) series.

%Fig_URhRuGe_TC_versus_doping

The evolution of magnetic order in correlated 4$f$- and
5$f$-electron metals is often discussed in terms of a simple
Doniach picture \cite{Doniach-PhysicaB-1977}, where the
competition between the on-site Kondo interaction and inter-site
RKKY (Ruderman-Kittel-Kasuya-Yosida) interaction determines the
ground state. In the Doniach model the control parameter is the
ratio of the exchange interaction $J$ over the bandwidth $W$.
Keeping $W$ constant, a weak hybridization ($J$ small) favours the
RKKY interaction and a magnetic ground state, while a strong
hybridization ($J$ large) favours a non-magnetic Kondo-screened
ground state. In the generic Doniach phase diagram the magnetic
ordering temperature $T_{\rm M}$ goes through a maximum with
increasing $J$ and vanishes when the RKKY and Kondo energies
become comparable. A typical example of a ferromagnetic material
that follows the generic Doniach phase diagram is cubic CeAg
($T_{\rm C}=5.6$~K)~\cite{Eiling-PRL-1981}. Here hydrostatic
pressure is used to control $J$ via the unit cell volume, such
that $T_{\rm C}(P)$ goes through a broad maximum near 0.7 GPa. For
the orthorhombic UTX alloys (X is Ge or Si) $J$ is not controlled
by the unit cell volume and the effects of hybridization are
difficult to control because of the strongly anisotropic magnetic
and electronic properties~\cite{Sechovsky-Handbook-1998}. For
instance, resistivity measurements under hydrostatic pressure on
pure URhGe show that $T_{C}$ increases linearly up to very high
pressures ($T_{C}$ reaches $\approx 17$~K at
13~GPa)~\cite{Hardy-PhysicaB-2005}, in disaccord with the Doniach
phase diagram.

Nevertheless, it is interesting to compare the volume effects due
to alloying and hydrostatic pressure. Assuming an isothermal
compressibility $\kappa = -V^{-1} ({\rm d}V/{\rm d}p)$ of
$\approx$0.8~Mbar$^{-1}$ \cite{Sakarya-PRB-2003}, substitution of
1 at.\% Ru, Co or Si leads to a chemical pressure of 0.37~kbar,
0.91~kbar or 0.46~kbar, respectively. Using d$T_{\rm C}$/d$p$ =
0.065~K/kbar as derived from the resistivity measurements under
pressure~\cite{Hardy-PhysicaB-2005} we calculate an increase of
$T_{\rm C}$ per at.\% Ru, Co or Si of 0.024, 0.059 and 0.030 K. In
the case of Ru and Co doping these calculated values are about a
factor 5 too small when compared to the measured initial increase
in $T_{\rm C}$, while in the case of Si doping $T_{\rm C}$ does
not increase at all. Clearly, chemical and mechanical pressure
give different results. Magnetization measurements under pressure
on URhGe doped with 0.325 at.\% Ru ($T_{\rm C}$ = 2.8 K at ambient
pressure) did not show a noticeable change of $T_{\rm C}$ for a
pressure of 4.3~kbar \cite{Sakarya-PhDThesis-2007}. This indicates
an additional complication, namely d$T_{\rm C}$/d$p$ varies with
doping concentration.

A striking difference between Co and Si doping on the one hand and
Ru doping on the other hand is that the first two substitutions
are isoelectronic, while the latter depletes the $d$-band. In a
simple model, extracting electrons from the $d$-band results in an
additional strengthening of the $f$-$d$ hybridization, which in
turn leads to a larger exchange parameter $J$, favoring the Kondo
interaction. Specific-heat measurements on URh$_{1-x}$Ru$_{x}$Ge
support this idea
\cite{Sakarya-PhDThesis-2007,Huy-tobepublished-2007}. The linear
coefficient of the electronic specific heat $\gamma$ increases as
a function of $x$ and reaches a maximum value near $x_{cr}=0.38$,
i.e. the critical concentration for the suppression of
ferromagnetic order. For isoelectronic Co and Si doping, the
(in)variance of $T_{\rm C}(x)$ is attributed predominantly to
anisotropic hybridization phenomena related to the anisotropic
variation of the unit cell parameters. It would be interesting to
investigate whether more sophisticated models, like the one
proposed by Sheng and Cooper \cite{Sheng-JAP-1985}, could explain
the observed behaviour of the magnetic ordering temperature. By
incorporating the change in the $f$-density spectral distribution
under pressure in LMTO band-structure calculations, these authors
could explain the observed maximum in the magnetic ordering
temperature for compounds like UTe.

The URh$_{1-x}$Ru$_x$Ge series deserves ample attention because it
might present one of the rare opportunities to investigate a
ferromagnetic quantum critical point in $f$ electron systems at
ambient pressure. Evidence for a ferromagnetic quantum critical
point is provided by measurements of the low-temperature
electronic specific heat $c(T)$
\cite{Sakarya-PhDThesis-2007,Huy-tobepublished-2007}. At the
critical Ru concentration $x_{cr} \approx 0.38$ a pronounced
non-Fermi liquid term $c \propto T\ln T$ is observed over a wide
temperature interval 0.5-9 K. Such a behaviour has been proposed
within the itinerant electron model for a ferromagnetic quantum
critical
point~\cite{Stewart-RMP-2001,vLöhneysen-archive-2006,Millis-PRB-1993}.

\section{Conclusions}

We have investigated the evolution of ferromagnetic order in the
correlated metal URhGe by substitution of Ru, Co and Si.
Magnetization measurements, and in case of Ru also resistivity
measurements, have been performed and the variation of the Curie
temperature has been extracted from the data. In the case of Ru
and Co doping, $T_{\rm C}$ goes through a maximum near $x=0.05$
for Ru doping and $x=0.60$ for Co doping. $T_{\rm C}$ vanishes
near $x=0.38$ for Ru doping and $x \approx 1.0$ for Co doping. Si
doping does not affect $T_{\rm C}$ (at least up to $x=0.20$). For
Ru, as well as for Co doping $T_{\rm C}(x)$ follows the generic
Doniach phase diagram, but anisotropic hybridization effects
hamper a quantitative analysis. Depletion of the $d$-band enhances
the suppression of magnetic ordering in the case of Ru doping. We
conclude that the alloy systems URh$_{1-x}$Ru$_x$Ge with $x_{cr}
\approx 0.38$ and URh$_{1-x}$Co$_x$Ge with $x_{cr} \approx 1.0$
are interesting candidates to investigate the ferromagnetic
instability.

\ack

S.S. acknowledges financial support of the Netherlands
Organization for Scientific Research (NWO), enabling a working
visit to the University Karlsruhe. This work was part of EC Cost
Action P16 (ECOM).

\end{document}